\documentclass[aps,prl,reprint,amsmath,amssymb,superscriptaddress,longbibliography]{revtex4-2}

\usepackage{graphicx}
\usepackage{amsmath}
\usepackage{amssymb}
\usepackage{amsthm}
\usepackage{comment}
\usepackage{mathtools}
\usepackage{tabularx}
\usepackage{xcolor}
\usepackage{xurl}
\usepackage{hyperref}
\hypersetup{
  breaklinks=true,
  colorlinks=true,
  citecolor=blue,
  linkcolor=black,
  urlcolor=blue
}

\newcommand{\ket}[1]{\left| #1 \right\rangle}

\newcommand{\bra}[1]{\left\langle #1 \right|}

\theoremstyle{definition}

\begin{document}

%TC:ignore
\title{{Framework for Robust Quantum Speedups in Practical Correlated Electronic Structure and Dynamics}}

\author{Jielun Chen}
\affiliation{Division of Physics, Mathematics, and Astronomy, California Institute of Technology, Pasadena, California 91125, USA} 
\affiliation{Institute for Quantum Information and Matter, Pasadena, California 91125, USA}
\author{Garnet Kin-Lic Chan}
\email{garnetc@caltech.edu}
\affiliation{Division of Chemistry and Chemical Engineering, California Institute of Technology, Pasadena, California 91125, USA}
\affiliation{Institute for Quantum Information and Matter, Pasadena, California 91125, USA}
\affiliation{Marcus Center for Theoretical Chemistry, Pasadena, California 91125, USA}

\begin{abstract}
Proposed quantum advantage in electronic structure has so far required significant fine-tuning to find problems where classical heuristics fail. We describe how to obtain robust quantum speedups for correlated electronic structure and dynamics precisely in the regime where widely used classical heuristics are most successful. 
% \chris{In this regime, leading classical heuristics scale linearly (or even independently) with system size but incur a high-degree polynomial cost in the correlation length $L_c$; our advantage is therefore with respect to $L_c$. We demonstrate a relative quantum speedup of up to $L_c^{18}$ in practically relevant settings.}
\end{abstract}
%TC:endignore

\maketitle

\emph{Introduction.}\quad The quantum simulation of electronic structure \cite{aspuru-guzik2005simulated, whitfield2011simulation,seeley2012bravyikitaev,toloui2013quantuma,wecker2014gatecount,peruzzo2014variational,hastings2014improving,poulin2014trotter,mcclean2014exploiting,babbush2015chemical,babbush2016exponentially,babbush2017exponentially,reiher2017elucidating, babbush2018low,babbush2019quantum,mcclean2020discontinuous,motta2021low,campbell2019random,berry2019qubitization,kivlichan2019phase,burg2021quantum,lee2021even,su2021fault,wan2022randomized,zini2023quantum,berry2024quantum,oumarou2024accelerating,rubin2024quantum,georges2025quantuma,koizumi2025faster,low2025fast} has sparked immense enthusiasm for quantum computing in many applications. At the same time, in many areas, classical heuristic quantum chemistry is nominally efficient (polynomial cost) and often accurate. While such heuristics cannot solve all problems, and the polynomial cost is also a practical barrier, the quest for quantum advantage often resembles a delicate task of fine-tuning to identify problems where classical heuristics are not yet successful. Since the power of classical heuristics is hard to prove and is still increasing,  this can be a challenging task. 

Here, we devise a framework to obtain substantial quantum speedup for correlated electronic structure (and dynamics) calculations, \emph{precisely in the regime where many classical heuristics are most successful}.
Our target will not be to solve the Schr\"odinger equation exactly, but to output quantities at the same level of approximation as the classical heuristic.
%. We use the same assumptions that many classical heuristics rely on for efficiency, and output the same quantity as the classical heuristic. 
While our framework potentially enables speedup of many methods, we provide a detailed analysis of two: the computation of excitation energies and dynamics via the $m$-exciton Bethe-Salpeter equation (BSE), and the computation of correlated ground-state energies described by $m$-fold coupled cluster excitations. Under the relevant assumptions, standard classical heuristic solutions in these methods can be made to scale \emph{linearly} with (or even independent of, for a crystal) the number of atoms, and thus we do not seek speedup with total system size. Instead, under the same assumptions, the quantum speedup is expressed in terms of the effective correlation length $L_c$, related to the interaction range $R_c$. While the speedup we achieve is, in principle, exponential in $m$, most practical applications do not require $m >3$, thus we focus on $m\leq 3$. Physically, the problems therefore correspond to simulating a few effective excitations, rather than an extensive number of electrons. In these cases, we obtain relative speedups to classical sparse techniques of up to ${O}(L_c^{18})$,
%for the realistic setting $m \leq 3$, 
or the 19th power in polynomial degree $d$ and 7th power in $R_c$. Within an excitation basis encoding that yields the correlated analog of exponential compression in simulating independent particles~\cite{babbush2023exponential,somma2025shadow,stroeks2024solving,chen2024quantum}, this enables relevant quantum simulations of correlated electrons in systems of thousands of atoms with moderate fault-tolerant resources.

\emph{General framework for speedup.}\quad We target electronic structure beyond mean-field theory (i.e., density functional theory or Hartree-Fock theory) in the regime where polynomial cost classical heuristics are accurate.
Such heuristics rely on two assumptions for accuracy and efficiency. In the ground state, they assume the electronic structure is described by small fluctuations around the mean-field Slater determinant, parametrized by operators in the manifold of $m$ particle-hole excitations for small $m$ (e.g., $m\leq 3$). A famous example is the coupled cluster method~\cite{shavitt2009many}, where $m \leq 3$ denotes the  widely used approximation of coupled cluster with up to triples excitations~\footnote{Typically the triples are treated with additional approximations~\cite{shavitt2009many}.}, often referred to as the gold standard of quantum chemistry~\cite{bartlett2007coupled}. 
In the context of excited states, the (generalized) Bethe-Salpeter equation for multiexcitons presents a related structure, but for fixed $m$, and the experimentally relevant case is for small $m$, e.g., $m \leq 3$. A second assumption is the locality of correlation. In the ground state, this means that well-separated particle-hole pairs do not contribute to the correlation energy, leading to coupled cluster methods with linear cost in system size~\cite{schutz2001low}. In excited states, the expression of local correlation is more subtle: while quasiparticles and quasiholes interact over long distances via the Coulomb interaction, the fluctuations of the quasiparticles and quasiholes are correlated over a shorter range. This has similarly been used to construct linear-scaling GW-BSE implementations~\cite{merkel2023linear}. The availability of linear scaling classical algorithms in both cases clearly distinguishes our problem from the conventional quantum simulation setting. 

In this practically relevant setting, we now describe a mechanism for quantum speedup.
%Under these assumptions, we now argue that 
Extracting physical quantities reduces to linear algebra computation, namely, solving eigenvalue problems, linear systems, and time evolution. In the $m$-exciton BSE and $m$-excitation CC equations, these involve $2m$-body linear operators acting in a space of dimension $\sim L^{2m}$ where $L$ is the system size. However (as we justify later) we can imbue them with additional structure, namely, 
%, involving $2m$-body linear operators which are 
geometric locality with range $R_c$. 
%These effective Hamiltonian operators can be derived from the underlying electronic structure Hamiltonian even though it does not itself have geometric locality due to the $1/r_{12}$ tail of the pairwise Coulomb interaction.
%, and they act in the Hilbert space of (up to) $2m$ quasiparticles. 
We can imagine performing linear algebra computation classically with a degree-$d$ polynomial of the sparse operator applied to a sparse initial state (for simplicity, imagine a single initial nonzero entry). For a general sparse linear operator, the nonzero entries of the vector then grow exponentially in $d$, but because of geometric locality, light cone arguments improve this to $\sim(dR_c)^{2mD} = L_c^{2mD}$ which is the $m$-excitation correlation volume $V_m$ ($D$ is the dimensionality of space, which can be chosen $=3$ for concreteness) and $L_c = dR_c$ is the effective correlation length. 
The total cost to multiply the linear operator $d$ times is thus the correlation volume $\times$ the operator sparsity $\times$ $d$. Note that, for this initial state, the total system size does not appear in the cost, and without additional structure, the dependence on correlation volume cannot be reduced.

% Hamiltonian with pair-wise interactions, which is geometrically local with interaction range by $R_c$  (the forms will be specified for the particular problems later). We can imagine solving these equations by implementing degree-$d$ polynomial of the Hamiltonian, and apply it to an initial state (for simplicity, imagine the initial state has a single non-zero entry). By lightcone arguments, the sparsity of state grows to $\sim(dR_c)^{2mD} = L_c^{2mD}$ which may be viewed as the $m$-particle/$m$-hole correlation volume ($D$ is the dimensionality of space, which can be chosen $=3$ for concreteness). The total cost to multiply for $d$ times is thus correlation volume $\times$ Hamiltonian sparsity $\times$ $d$. Without non-trivial assumptions, the dependence on correlation volume cannot be significantly reduced.

%\rem{write in a more technical way} 

%Suppose the linear operator can be constructed as a block-encoding (a submatrix of a unitary), then the cost to encode a polynomial of it is merely $d$ times the block-encoding cost. The remaining cost comes from the success probability of post-selection, which can be bounded by the operator sparsity when using sparse block-encoding schemes. 

On the other hand, a quantum implementation of the above does not need to store the light cone data. For example, using the quantum singular value transform (QSVT) \cite{gilyen2019quantum}, and denoting the effective Hamiltonian by ${A}$, 
%a unifying framework for polynomial transformation of linear operators. QSVT 
QSVT operates on a block encoding of $A/\alpha$, where $\alpha$ ensures $\|A\|_2/\alpha \leq 1$, and using sparse block-encoding schemes, 
%$\alpha = $ operator sparsity $\times$ maximal element. For polynomials in our application, 
the total cost of QSVT to apply the polynomial is the block-encoding cost $\times$ $\alpha$ $\times$ $d$ (other overheads 
%with post-selection and measurement 
can enter depending on the property to be computed, which we describe later). Since the operator sparsity and polynomial degree cancel between the quantum and classical costs, and assuming matrix elements are bounded by $O(1)$ \footnote{Technically, the matrix elements are a summation of poly($m$) terms with magnitude $O(1)$, but throughout the paper we focus on $m \leq 3$ so we drop this factor.}, this gives a source of speedup,
\begin{equation}
    \text{quantum speedup} = \frac{\text{correlation volume}}{\text{block-encoding cost}}. \label{eq:speedup}
\end{equation}
% Physically, the block-encoding cost comes only from specifying the data for the linear operator that generates the correlations, 
% not the correlation volume itself. 
Depending on the input model, the amount of data 
ranges from $O(1)$ (specifying the geometry of a parametrically defined system such as a crystal) to 
$\sim V L_c^{2D}$ (specifying Coulomb integrals in a non-translationally invariant problem of volume $V$)
but is never exponential in $m$, 
% parameterized systems such as crystals) is only required to $\sim m^2L_c^{2D}$ (pair-wise Coulomb) or more, but is never exponential in $m$. 
which forms the basis of high-order polynomial speedups for fixed $m$. %In Fig.~\ref{fig:volume}, we illustrate the source of speedup through a simple example: a particle-hole pair on a 1D lattice. 
In the remaining part, we always work in the realistic setting of $m \leq 3$.

%\begin{figure}
%    \centering
%    \includegraphics[width=0.9\linewidth]{volume.png}
%    \caption{(a) A particle-hole pair on a 1D lattice (i.e. $m=1,D=1$). (b) The classical and quantum cost of computing the polynomial transformation of the effective Hamiltonian. With interaction range $R_c$, the sparsity of the operator is $O(R_c)$ in 1D. The correlation volume is an area of radius $L_c=dR_c$ on an effective $2$-dimensional lattice, with coordinates denoting orbital indices of the particle and hole. Classical sparse algorithms must keep track of $O(L_c^2)$ elements, and total computation takes $O(L_c^2) \times O(R_c) \times O(d)  = O(L_c^3)$ time. On the other hand, if block-encoding takes $O(1)$ time, quantum algorithms take $O(L_c)$ time to encode the entire volume, a cubic speedup. In the text, $D=3$ and $m > 1$, leading to larger polynomial speedups.}
%    \label{fig:volume}
%\end{figure}

% describe the framework that enables this speedup and give a more detailed analysis in two specific problem settings, the $m$-particle/$m$-hole Bethe-Salpeter equation (BSE) and linearized coupled cluster (CC) with up to $m$-excitations, for the realistic cases of $m \leq 3$.

\emph{Problem setup.} \quad % {We will be devising quantum algorithms for post-mean-field electronic structure to output the result of the classical post-mean-field algorithm with speedup. 
We assume the standard electronic structure problem, specified by the electronic structure Hamiltonian ${H}$ for $n_e$ particles in an external potential, and a finite single-particle basis (total basis dimension $L\sim \mathrm{const}\times V$, where $V$ is the physical volume)~\cite{szabo1996modern}. We will be interested in the eigenstates and dynamics generated by $H$. Additionally, similar to the classical post-mean-field algorithms, we assume
(1) The mean-field calculation has been performed to obtain a determinant of molecular orbitals $|0\rangle$, and a set of exponentially localized (with radius $R_\text{loc}$) molecular orbitals exist and have been obtained classically. 
% (2)
%  There is a finite correlation length $L_c$. (3) 
%      The target state can be expressed with excitation operators parametrized by the space of up to $m$-particle-hole-excitations with small $m$. 
     (2) A simple initial guess serves as a good initial state. %\rem{merge with previous paragraphs?}

\emph{Excitation space representation.}\quad Much of quantum simulation is formulated in the second-quantized occupancy representation, where each qubit represents a single orbital. In contrast, the linear algebra in the post-mean-field electronic structure algorithms takes place in the Hilbert space of $m$-fold excitations, also referred to as the configuration interaction or first quantized representation in the literature~\cite{babbush2017exponentially,su2021fault} (for the distinction, see Ref.~\cite{su2021fault} and Sec.~II.C of Supplemental Material~\cite{supplemental}).
\nocite{camps2022fable,camps2023explicit,sunderhauf2024blockencoding,babbush2018encoding,giovannetti2008quantum,aaronson2015read,arunachalam2015robustness,matteo2020faulttolerant,hann2021resilience,chen2023efficient,lin_lecture_2022,berry2015hamiltonian,helgaker2000atomic,helgaker2000gaussian,obara1986efficient,helgaker2000molecular,sharma2021efficient,low2024trading} 

Concretely, consider the Hartree-Fock state as a Fermi vacuum $|0\rangle$. Hole (occupied) creation operators will be labeled by subscripts $i, j, k, \ldots$, particle (virtual) creation operators by subscripts $a, b, c, \ldots$, and general orbitals by $p, q, r, s$; these states are all localized as assumed above. We then have one-particle and one-hole (single excitations) $c^\dagger_a c_i |0\rangle \equiv |{}_i^a\rangle$, two-particle and two-hole, etc., up to $m$-particle and $m$-hole ($m$-fold excitations) states. The $m$-fold excitation is uniquely labeled by an excitation string $|\mu\rangle \equiv |{}_{ijk\ldots}^{abc\ldots}\rangle$, with $i>j>k \ldots; a>b>c \ldots$. Because of the sorted indices, the multiexcitation space is not a tensor product of single-excitation spaces. 

%\rem{One thing to note about antisymmetrizer in \cite{berry2018improved} is that they have to apply on an ordered string to generate the anti-symmetric states; if the states are not ordered, the ancilla will not be set back to $0$.}
% If we consider a larger basis where the particles/holes are distinguishable, then we can identify the $m$-fold exciton basis with antisymmetrized states of $m$-holes and $m$-particles. (Later we briefly consider the use of the distinguishable basis in the context of state preparation, where the basis corresponds to a first quantization basis).

The total fermionic Hilbert space for $n_e$ particles is spanned by $\{|0\rangle, |{}_i^a\rangle, |{}_{ij}^{ab}\rangle \ldots\}$. We can consider a qubit encoding where $\mu$ is replaced by its binary encoding, using $O(2m \log L)$ qubits for the $m$-fold excitations. If $n_e \propto L$ and if we seek an exact solution where $m \propto L$, then there is little saving versus the   second quantized representation, thus prior quantum simulation work in the excitation basis has focused instead on small $n_e$ (i.e., small physical systems). However, we focus on small $m$ (as used in classical heuristics) which allows us to efficiently encode the state of large systems. For example, for a few thousand atoms, with a reasonable atomic basis, we might take $L=10^5$, requiring $33m$ qubits, and for $m=3$ this is only 99 logical qubits for the state. This is to be compared to $152$ qubits to store the 76 orbital active space second-quantized representation of the 18-atom FeMo cofactor. This exponential compression is  analogous to that in  works on simulating exponentially many non-interacting fermions or oscillators~\cite{babbush2023exponential,somma2025shadow,stroeks2024solving,chen2024quantum}, but in the few quasiparticle excitation setting. 

\emph{Locality and connectedness.} \quad We emphasize that local orbitals alone do not make excitation-space matrices sparse, due to Coulomb tails and vacuum couplings; see Sec.~I.A of Supplemental Material~\cite{supplemental} for further explanation. We therefore restrict either to the fixed-$m$ multiexciton BSE, or work with the Liouvillian $L=[H,\cdot]$, whose commutator structure preserves geometric locality and appears in the (linearized) coupled cluster equations.

\emph{Multiexciton energies and dynamics.}\quad Optoelectronic properties of semiconductors underpin much of modern technology. 
% across a myriad of sectors in our lives: display technology, energy harvesting, lasers and telecommunication. 
The basic unit of a photoexcitation is the exciton ($m=1$), a quasiparticle-quasihole pair. While single excitons are arguably well described by existing classical techniques, in a large class of semiconductors, multiexciton states are important, particularly under higher flux conditions in lasing applications. For example, triexcitons ($m=3$)  have been studied experimentally in nanocrystalline CdSe~\cite{shulenberger2021resolving} and multiexcitons have been invoked as behind carrier multiplication in bulk crystalline PbS and PbSe~\cite{pijpers2009assessment}. 

Despite their importance, it has not been possible to model multiexcitons in a first-principles atomistic manner. The starting point 
% would be to compute the $m$-exciton wavefunctions and energies, from 
is the $m$-exciton BSE equation~\cite{steinhoff2018biexciton,kuhn2019combined}. In the Tamm-Dancoff approximation~\cite{fetter2012quantum}, and using static screening and self-energies~\cite{blase2020bethe}, the exciton energies and dynamics are obtained from 
%the eigenvalue problem 
$\mathbf{A} \mathbf{c} = e \mathbf{c}$ and $i \partial_t \mathbf{c} = \mathbf{A} \mathbf{c}$ respectively, 
where $\mathbf{A}$ is the multiexciton effective Hamiltonian in the $m$-fold excitation basis, $\mathbf{c}$ is the amplitude of the multiexciton wave function, and $e$ is the energy. 
Further details are given in Sec.~I.B of Supplemental Material~\cite{supplemental}.

% As discussed above, the restriction to the $m$-excitation manifold means that $\mathbf{A}$ does not contain contributions from vacuum-like fluctuations. 
We now argue the Coulomb interaction does not prevent sparsity of $\mathbf{A}$ (in realistic systems). This is because the nonlocality of the Coulomb interaction is ``diagonal'', and the decay in the separation of exciton-exciton centers (which determines the off-diagonal sparsity of $\mathbf{A}$) goes like $1/r_{12}^3$ (where $r_{12}$ is the exciton-exciton center separation).
Thus, the $\mathbf{A}$ matrix is heavily concentrated around the diagonal, and classical calculations take advantage of this sparsity to truncate small terms and obtain linear scaling with system size~\cite{merkel2023linear}. 

Given some linear range $R_c$ in the off-diagonal elements, then in $D$ dimensions, the off-diagonal scattering of each particle or hole involves a volume $R_c^D$. The Hamiltonian's sparsity is the union of single-particle volumes, $s \sim 2mR_c^D$ (assuming matrix elements where more than one quasiparticle changes state are not dominant). With this we can estimate the classical cost. We first consider the eigenvalue problem. As outlined above, $d$ iterations of a classical iterative eigensolver applied to an initial unit vector have cost $\sim (dR_c)^{2m D} \times R_c^{D} \times d$~\footnote{Assuming quasi-particle hops, the factor $(dR_c)^{2mD}$ can be computed by estimating the number of sites reached by a walker on a $2mD$ dimensional lattice after $d$ steps.}. 
{ We determine $d$ as the number of iterations to 
% We might assume that if there is a threshold on the vector elements, the sparsity of the vector stops growing by some $d$, before the non-zero values are fully converged by the eigensolver. Denoting $d'$ as the number of iterations to fully 
converge the eigenvector to a given element threshold.} For computing the lowest $m$-exciton given a trial state of overlap $\gamma$, a typical scaling is $d=O(\Delta^{-1}\log(\epsilon^{-1}\gamma^{-1}))$ \cite{saad1980rates, saad1992numerical}, where $\Delta$ is the gap and $\epsilon$ is the two-norm state error, for simplicity.
%, we  assume $d' \propto d$. 
The above initial state is appropriate if the center of mass of the $m$-exciton is localized (although its extent in the relative quasiparticle coordinates can be large); if the center of mass is fully delocalized, one can expect an additional $V$ factor in the cost. 

%{If we assume an initial state already delocalized across the correlation volume, i.e. with $(dR_c)^{2m D}$ matrix elements, the above costs are multiplied by a constant factor.}

% % We can hence estimate the cost of $d$ iterations of a classical iterative eigensolver. A degree $d$ application will fill-in the vector up to $V_m \sim (dR_c)^{2m D}/(2mD)!$ elements and the total number of single particle states touched (which determines the total number of integrals we need in the calculation) is $V_1 \sim 2m (dR_c)^D$, thus the total cost is $(dR_c)^{2m D}/(2mD)! \times 2m R_c^{D} \times d$. 

% For ground state preparation with a trial state of overlap $\gamma$, a typical scaling is $d=O(\Delta^{-1}\log(\epsilon^{-1}\gamma^{-1}))$ \cite{saad1980rates, saad1992numerical}, where $\Delta$ is the gap and $\epsilon$ is the 2-norm error of the state.

We now outline the quantum calculation under the same sparsity assumptions for different input types. With the block-encoding, we can use several quantum linear algebra techniques to obtain the eigenvalues; we use phase estimation \cite{kitaev1995quantummeasurementsabelianstabilizer, somma2019quantum, obrien2019quantuma, dutkiewicz2022heisenberglimited, ding2023simultaneous, li2023adaptive, wang2023quantum} as an example,  which has cost $\sim (C_\text{init} + C_\text{BE} sd')/\gamma$. Here, $C_\text{BE}$ is the cost to construct the block-encoding, $C_\text{init}$ is the state preparation cost, $s$ is the sparsity, and $d'$ is the polynomial degree and may be viewed as the total evolution time. [In phase estimation we may choose not to prepare any accurate state (beyond initial state preparation) and then $d'$ reflects the readout cost $\sim (1/\epsilon)$. For given $\epsilon, \gamma$, we simply here take the quantum $d' \propto$ classical $d$]. 

We briefly outline details of $C_\text{init}$ and $C_\text{BE}$. If we are interested in an $m$-exciton in a local region of space, we may assume $1/\gamma$ is acceptable even when using an initial unit vector. If a more delocalized state is required, we may prepare the Hartree-Fock state, as typically done in the classical calculation. To do so, it is easiest to work in the distinguishable particle excitation basis (see Secs.~II.A and II.C of Supplemental Material~\cite{supplemental}), 
%which is achieved by
% . This means that the ordering $i>j>k$ etc., is not imposed, and the state is a product space. Then, the Hartree-Fock excited state can be prepared by 
%applying rotations together with antisymmetrization, 
following the procedure in Ref.~\cite{berry2018improved}, which contributes negligible cost.
For $C_\text{BE}$, we consider three relevant input models: loading all integral information; only atomic information; and atomic information in a translationally invariant problem (crystals). The corresponding costs {and input models} are analyzed in Sec.~II.E of Supplemental Material~\cite{supplemental} in detail, and the resulting speedups are shown in Table~\ref{tab:speedups}.
The QROM and QRAM cost models underlying this analysis are discussed in Sec.~II.B of Supplemental Material~\cite{supplemental}.

The above discussion has focused on the starting point of  $m$-exciton energies.
%We have so far described computing the $m$-exciton eigenstates. 
However, as in standard quantum simulations, the dependence on initial state overlap $\gamma$ complicates discussions of asymptotic speedup~\cite{lee2023evaluating}. { A cleaner picture of advantage may be provided by the $m$-exciton dynamics which are also described by the operator $\mathbf{A}$ (see Ref.~\cite{steinhoff2018biexciton}).}
The initial-state overlap and preparation assumptions are discussed further in Sec.~II.D of Supplemental Material~\cite{supplemental}.

% The time-dependent electronic structure setting may hold additional advantages when comparing to alternative classical simulation techniques, as we discuss briefly later.

\begin{table*}[t!]
    \centering
    \begin{tabular}{|c||c|c|c||c|c|c|}
        \hline
         & \multicolumn{3}{c||}{\text{Multi-exciton BSE ($m=3$)}} & \multicolumn{3}{c|}{\text{Linearized coupled cluster ($m=3$)}}  \\
         \hline
         Input model  & Quantum cost & Speedup (ratio) & Speedup (power) & Quantum cost & Speedup (ratio) & Speedup (power)\\
        \hline
        Integrals & $d^{2D+1}R_c^{3D}$ & $L_c^{12}$ & $(2.7,2.3)$ & $VdR_c^{3D}$ & $d^{6}L_c^{12}$ & $(3.2,2.3)$ \\
        \hline
        Atomic data & $d^{D+1}R_c^{2D}$ & $L_c^{15}$ & $(4.8,3.5)$ & $VdR_c^{2D}$ & $d^{3}L_c^{15}$ & $(6.3,3.5)$\\
        \hline
        Crystals & $d R_c^{D}$ & $L_c^{18}$ & $(19,7)$  & $dR_c^{2D}$ & $d^{3}L_c^{15}$ & $(6.3,3.5)$ \\
        \hline
    \end{tabular}
    \caption{Summary of speedups for different input models compared to the dominant classical cost $d^{2mD+1}R_c^{(2m+1)D}$ (or $V d^{2mD+1}R_c^{(2m+1)D}$ for noncrystals in LCC). The exciton case assumes the center of mass is not delocalized. The speedup is compared for fixed $m=3, D=3$, and is given in two ways: (1) the ratio of classical cost and quantum cost, (2) we express the classical dependence on $d$, $R_c$ individually as powers of the classical cost [e.g., if the classical cost $d^{a_1}R_c^{b_1}$ and the quantum cost is $d^{a_2} R_c^{b_2}$, we report $(a_1/a_2, b_1/b_2)$]. For the problem of multiexciton eigenvalue determination, one should include $1/\gamma$ in the quantum cost factor, where $\gamma$ is the initial state overlap.}
    \label{tab:speedups}
\end{table*}

%Integrals & $d^{2D+1}R_c^{3D}/\gamma$ & $L_c^{(2m-2)D}$ & $(2.7,2.3)$ & $VdR_c^{3D}/\epsilon$ & $d^{2D}L_c^{(2m-2)D}$ & $(3.2,2.3)$ \\
%Atomic data & $d^{D+1}R_c^{2D}/\gamma$ & $L_c^{(2m-1)D}$ & $(4.8,3.5)$ & $VdR_c^{2D}/\epsilon$ & $d^{D}L_c^{(2m-1)D}$ & $(6.3,3.5)$\\
%Crystals & $d R_c^{D}/\gamma$ & $L_c^{2mD}$ & $(19,7)$  & $dR_c^{2D}/\epsilon$ & $d^{D}L_c^{(2m-1)D}$ & $(6.3,3.5)$ \\

\emph{Linearized coupled cluster theory.}\quad Another common post-mean-field electronic structure task is to obtain accurate ground-state energetics. 
In the widely applied coupled cluster method, it is well established that up to triples ($m=3$) excitations are required for accuracy. While linear scaling implementations of coupled cluster with up to triples  exist~\cite{schutz2001low,guo2018communication}, it is challenging to converge them with respect to the cutoff $R_c$ in small bandgap problems with a high atomic density (such as in the common inorganic semiconductors discussed for the multiexciton problem). As an application of our framework, we consider the linearized version of coupled cluster theory (LCC)~\cite{taube2009rethinking}, which is generally faithful to the full coupled cluster theory where coupled cluster theory is accurate, i.e., where the $\hat{T}$ amplitude is small.

% wide
% A second potential use case in the area of semiconductors is to compute accurate ground-state energetics. Indeed, as semiconductors incorporate new elements and new morphologies and structures, the energetics is relevant to understand material phase separation, impurity distribution, and mechanical properties. Since it is well established that 

% triples excitations are required for reasonable accuracy, and one also requires an extensive theory, this is a natural application of (linearized) coupled cluster singles, doubles, and triples (CCSDT)~\cite{shavitt2009many}. (The small bandgaps of many semiconductors suggest we avoid a simpler perturbative treatment of triples; in addition, the linearized version of coupled cluster theory is generally faithful to the full coupled cluster theory in the region where coupled cluster theory is accurate, i.e. where the $\hat{T}$ amplitude is small). 

The linearized coupled cluster ansatz is
\begin{align}
    |\Psi\rangle = (1+ {T})|0\rangle
\end{align}
 where 
 %$|0\rangle$ is the Fermi vacuum and 
 ${T} = T_1 + T_2 + \ldots = \sum_\mu t_\mu {O}_\mu$, and ${O}_\mu$ excites to determinant $|\mu\rangle$, and $T_m$ contains the subset in the $\mu$ sum corresponding to $m$-fold excitations.
% The linearization refers to the approximation of the standard coupled cluster operator $e^{\hat{T}}\approx (1+\hat{T})$ which applies in the regime where coupled cluster theory is accurate (i.e. $||\hat{T}||$ is small). 
The LCC amplitude equations are
\begin{equation}
        \sum_{\nu} t_{\nu} \bra{\mu} [{H},{O}_\nu]\ket{0} = -\bra{\mu}{H}\ket{0}
    \end{equation}
Further details of the linearized coupled cluster equations are given in Sec.~I.C of Supplemental Material~\cite{supplemental}.
% where $t_\nu$ corresponds to the cluster excitation amplitude and $\hat{O}_\nu$ is an excitation operator to the $|\mu\rangle$ excited determinant. 
In the case of LCCm, the maximum excitation is to an $m$-particle and $m$-hole state. The ground-state correlation energy is computed as $    E_c = \langle 0 |{H} {T}_2 |0\rangle$
where we have assumed Brillouin's theorem~\cite{szabo1996modern} (the ${T}_1$ contribution vanishes within the linearized CC ansatz). Unlike for truncated configuration interaction, the correlation energy of truncated (fixed $m$) linearized coupled cluster theory is extensive, i.e., $\lim_{V\to \infty} E_c/V = \textrm{const}$. \footnote{For completeness, we note that the term linearized CC theory has to our knowledge appeared once before in the context of quantum simulation~\cite{baskaran2023adapting}, but the equations used in that work contain an important error as they mistakenly replace the commutator $[\hat{H}, \hat{O}_\nu]$ by $(\hat{H}-E)\hat{O}_\nu$ (reducing the problem to a linear equation in the Hamiltonian, rather than the Liouvillian), which is not LCC. Further the work in Ref.~\cite{baskaran2023adapting} which focuses on adaptations to NISQ and early fault tolerance, does not identify the characteristics for quantum advantage discussed here.}\nocite{baskaran2023adapting}

We now confirm that the long-range Coulomb does not eliminate the Liouvillian sparsity within standard assumptions of classical heuristics. 
Matrix elements of the Hamiltonian within a fixed $m$-fold excitation basis have been discussed above. In LCC, however, the Hilbert space is expanded to include different $m$ states, e.g., for $m=2$, the excitation basis is $\{ |0\rangle, |{}_i^a\rangle, |{}_{ij}^{ab}\rangle\}$, and there are new matrix elements $\langle 0 | {H} |{}_i^a\rangle$, $\langle 0|{H} | {}_{ij}^{ab}\rangle$, $\langle {}_i^a | {H} | {}_{kl}^{cd}\rangle$.
%interested in $\{ |0\rangle, |{}_i^a\rangle, |{}_{ij}^{ab}\rangle\}$. The fact that we have determinants with different levels of excitations introduces new matrix elements compared to the $m$-exciton problem before. 
Also in coupled cluster theory, the interaction is not screened. 
% To take advantage of locality we again consider the decay of the matrix elements. 
We can identify that the additional matrix elements decay like charge-dipole and dipole-dipole interactions. 
% (For example, one can verify that the $\langle 0|\hat{H}|S\rangle$ contains a term that decays like a charge-dipole interaction and $\langle 0|\hat{H}|D\rangle$ decays like a dipole-dipole interaction). 
In classical local correlation, it is then argued that because we work in a weak perturbation regime (as a condition for coupled cluster theory to be accurate), in perturbation theory the energy contributions enter at 2nd order in dipole-dipole interactions {(with the charge-dipole contribution entering at {4th} order if Brillouin's theorem is satisfied)}. We can therefore neglect such Hamiltonian matrix elements beyond some radius 
%thus if we cut off off-diagonal Liouvillian elements with some radius $
$R_c$, with each neglected energy contribution {up to 3rd order in the interaction} being $O(1/R_c^6)$. By the same argument, the Liouvillian action based on these matrix elements is sparse and geometrically local with radius $O(R_c)$.

The classical algorithm to solve the amplitude equations is discussed in standard works~\cite{saebo1993local,schutz2001low} and we give a simplified presentation.  The initial guess is usually the second-order perturbation theory amplitude, for which approximate (semi-canonical) estimates can be obtained with a cost comparable to the classical mean-field~\cite{pinski2015sparse}. 
% The cheapest initial guess is the semi-canonical MP2 amplitude, which encodes the doubles $\hat{T}_2$ elements as $t_{ijab} = \langle ij||ab\rangle /\Delta_{ijab}$ where $\Delta_{ijab}=\epsilon_i + \epsilon_j -\epsilon_a - \epsilon_b$, i.e. differences of the Fock operator diagonal in the semi-canonical basis. Given a problem with physical volume of $V$, the semi-canonical MP2 amplitude can be computed classically with $O(V^4)$ cost; using the sparsity arguments in the local basis, it can be computed with $O(V)$ cost. 
The initial doubles amplitudes contain $O(V R_c^D)$ nonzero elements. Two physical processes then fill in the vector during the iterations to solve the equations; scattering between different $m$-fold excitation manifolds, and scattering within a given $m$-fold excitation manifold, and the rate of fill in depends on the problem.  
If we assume that the $T$ amplitude sparsity is dominated by the amplitudes in the highest $m$th order excitation and the fill in is generated by single electron scattering (i.e., one-electron and charge-dipole two-electron matrix elements), the sparsity follows the $m$-exciton expression, $s\sim{2mR_c^{D}}$, and the correlation volume scales as $O(VL_c^{2mD})$. 

For quantum calculations, similar to the exciton case, given the block encoding we can choose one of many quantum linear system solvers \cite{harrow2009quantum, ambainis2010variable, childs2017quantum, gilyen2019quantum, subasi2019quantum,lin2020optimalb,costa2021optimala,an2022quantum,dalzell2024shortcut,jennings2025randomizeda, morales2025quantumlinearsolverssurvey} to output a state proportional to the solution, where {the current optimal scaling of the query complexity is $O(s d)$, where $d=O(\kappa \log(1/\epsilon))$ with $\kappa$ being the condition number of the linear operator}. The total quantum cost is $\sim (C_\text{init} + C_\text{BE}sd)/\epsilon'$, where $\epsilon'$ is the error from measurement, discussed further below.
%we describe below how to translate this into the energy error). 
For the initial state, if we choose to directly load classically computed initial amplitudes, various state preparation techniques use {$C_\text{init}\sim VR_c^D$ gates} \cite{grover2000synthesis,grover2002creating,plesch2011quantumstate,sanders2019,zhang2022quantum,sun2023asymptotically,yuan2023optimal,gui2024spacetimeefficient,lemieux2024quantum,mcardle2025quantumstatepreparationcoherent}. Alternatively, in a crystal, one can write a function that takes only local amplitude data and
once this local information is loaded, 
%one can 
%prepare local amplitudes through state preparation, and 
use translation-invariance to ``fill out'' the remaining entries of the double amplitudes with {polylog} additional arithmetic operations, yielding $C_\text{init}$ independent of $V$. % \sim \mathrm{const}$.
%can be viewed as constant}. 
The corresponding costs for $C_\text{BE}$ under the assumptions of loading integrals, atomic data, or for crystals, are described in Sec.~II.E of Supplemental Material~\cite{supplemental}.

 We aim to estimate $E_c/V$ to an additive error $\epsilon$ where $E_c=\langle 0| {H} {T}_2|0\rangle$. This can be achieved by estimation of an overlap (dot product) between the  $T$-amplitudes and the Hamiltonian integrals. The latter state can be prepared with $O(VR_c^D)$ gates. Both states normalize the original vectors by $O(\sqrt{VR_c^D})$, thus on a quantum computer we need to resolve the states to accuracy $\epsilon'=\epsilon/R_c^D$, incurring an $O(R_c^D/\epsilon)$ overhead via amplitude estimation and improved variants \cite{brassard2002quantum, aaronson2019quantum,grinko2021iterative,fukuzawa2023modified,rall2023amplitude,harrow2019adaptive,cornelissen2023sublineartime,dalzell2023quantum}. We drop the dependence on $\epsilon$ in the cost as we target constant precision. The total speedups are summarized in Table~\ref{tab:speedups}.

\emph{Discussion.}\quad We have presented a framework for quantum speedups of post-mean-field electronic structure algorithms. Our speedup does not require fine-tuning or assuming a worst-case classical exponential complexity with system size, but instead retains the standard classical heuristic assumptions of locality and low-excitation rank in the fluctuations from mean-field. We present a simple mechanism for quantum speedup,
where the quantum algorithm eliminates the need to track correlations within a given correlation volume of $2m$ quasiparticles and holes. We presented examples of excited states, dynamics,  and ground-state computation. Out of these, we consider the dynamics of multiexcitons to be most promising, both due to the larger speedups, and because there are few alternatives to solving the Schr\"odinger equation for this problem. For example, the dynamical formation and breakup of high-energy multiexcitons resemble other scattering processes known to be challenging for classical methods~\cite{farrell2025digital}. Further, reaching the high-energy regime for a few quasiparticles (as studied here) is more accessible in real materials than the high-energy regime for an extensive number of electrons, increasing the relevance of such challenging dynamics. On the other hand, the linearized coupled cluster approach discussed here also has unique properties as a ground-state algorithm: for example, it eliminates dependence on the initial overlap.

% Because we allow for the typical assumptions that make classical heuristics successful, and because of the simple mechanism for quantum speedup, we consider this to be a robust strategy to find quantum speedups across many different existing electronic structure methods. At the same time, 
In evaluating quantum speedup, we have compared to classical sparse linear algebra, the foundation of the most common reduced-scaling classical electronic structure methods. The classical sparsity has been evaluated on asymptotic expressions and in further work should be quantified precisely for specific systems. Although sparsity is incompressible in the worst case, we may wish to evaluate other strategies that could help to reduce the classical cost in specific applications, for example, stochastic techniques or rank reduction, such as tensor hypercontraction~\cite{shenvi2013tensor} or quantics tensor trains (QTT)~\cite{kuhn2019combined}. 
Each of these must be assessed on a case-by-case basis (see Sec.~III of Supplemental Material~\cite{supplemental}), and further classical cost reduction will require problem structure beyond that considered here. 
For example, the exponential dependence on $m$ and $D$ is replaced in the QTT representation by the dependence on the bond dimension, which is governed by the state ``smoothness'' as well as the entanglement in the particle representation, and these will be different for a low-energy $m$-exciton stationary state and entanglement-generating high-energy $m$-exciton dynamics.  The large polynomial separations we see, e.g., for crystal triexciton calculations, provide optimism that 
a practical quantum versus classical speedup cannot be completely removed. {Further, extensions beyond electronic problems, for example to include phonon and photon coupling, can be pursued~\cite{white2020coupled,haugland2020coupled}.} {Also, the mechanism by which speedup is achieved, namely for large correlation length in $m$-quasiparticle simulations, may enable speedup of other types of electronic structure and dynamics heuristics.} In summary, our framework provides a new avenue to pursue quantum speedups to study correlated electronic structure and dynamics at the largest scales at which quantum simulations are in demand today.

%TC:ignore
\emph{Acknowledgments.}\quad We thank Alex Dalzell, Jiaqing Jiang, Jiace Sun, Lin Lin, Sam McArdle, and Samson Wang for helpful discussions and comments. We acknowledge support from the US Department of Energy, Office of Science, under Grant No. DE-SC0019374.


%TC:endignore

\clearpage
\onecolumngrid
\setcounter{page}{1}
\renewcommand{\thepage}{S\arabic{page}}
\setcounter{equation}{0}
\setcounter{table}{0}
\setcounter{figure}{0}
\setcounter{section}{0}
\makeatletter
\renewcommand{\theHequation}{SI.\arabic{equation}}
\renewcommand{\theHtable}{SI.\arabic{table}}
\renewcommand{\theHfigure}{SI.\arabic{figure}}
\renewcommand{\theHsection}{SI.\arabic{section}}
\@ifundefined{theHfrontmatter}{}{\renewcommand{\theHfrontmatter}{SI.\arabic{frontmatter}}}
\makeatother

\begin{center}
{\bfseries Supplementary information for ``Framework for robust quantum speedups in practical correlated electronic structure and dynamics''\par}
\vspace{0.7em}
Jielun Chen$^{1,2}$ and Garnet Kin-Lic Chan$^{3,2,4}$\\
\vspace{0.35em}
{\small
$^{1}$Division of Physics, Mathematics, and Astronomy, California Institute of Technology, Pasadena CA 91125, USA\\
$^{2}$Institute for Quantum Information and Matter, Pasadena CA 91125, USA\\
$^{3}$Division of Chemistry and Chemical Engineering, California Institute of Technology, Pasadena CA 91125, USA\\
$^{4}$Marcus Center for Theoretical Chemistry, Pasadena CA 91125, USA\\
\texttt{garnetc@caltech.edu}}
\end{center}
\vspace{1em}

\section{Effective Hamiltonians}
We present more details of the effective Hamiltonians in the multi-exciton and LCC setting. They will be derived based on the ab initio Hamiltonian
\begin{equation}
H = \sum_{pq} t_{pq} c_p^\dagger c_q +\frac{1}{2} \sum_{pqrs} V_{pqrs} c_p^\dagger c_q^\dagger c_s c_r 
\end{equation}
where $c, c^\dag$ are electron annihilation and creation operators for spin orbitals{, and the 1-electron integral is
\begin{equation}
t_{pq} = \int d\mathbf{r} \, \varphi_p^*(\mathbf{r}) \left( -\frac{1}{2} \nabla^2 - \sum_{A} \frac{Z_A}{|\mathbf{r} - \mathbf{R}_A|} \right) \varphi_q(\mathbf{r})
\end{equation}
and the 2-electron integral uses the physicists' index order convention
\begin{equation}
    V_{pqrs} =\int d\mathbf{r}_1 \int d\mathbf{r}_2 \frac{\varphi_p^*(\mathbf{r}_1) \varphi_q^*(\mathbf{r}_2) \varphi_r(\mathbf{r}_1) \varphi_s(\mathbf{r}_2)}{|\mathbf{r}_1 - \mathbf{r}_2|}
\end{equation}
where we assume $\varphi_p $'s are localized molecular orbitals}. The Fock operator $f_{pq}$ takes the form
\begin{equation}
f_{pq} = t_{pq} + \sum_{i\in {\text{occ}}} \left( V_{piqi} - V_{piiq} \right)
\end{equation}
where occ refers to the occupied space with respect to the Hartree-Fock ground state $\ket{0}$. {In the basis of localized molecular orbitals, $f$ is block-diagonal with respect to the occupied and virtual blocks, i.e. $f_{pq}=f_{qp}=0$ if $p$ is in the occupied block and $q$ is in the virtual block.}

{
\subsection{Locality, sparsity, and connectedness} One might assume working in the local molecular orbital basis immediately leads to geometric locality and sparse matrices in the linear algebra, but the picture is complicated in the \textit{ab initio} setting by the long-range Coulomb interaction and the projection into the excitation basis.
%Denoting a general orbital (either occupied or virtual) by labels $p, q, r, s$, 
The electronic Hamiltonian is parametrized by the one-electron and two-electron integrals with elements $t_{pq}$, $V_{pqrs}$.
%$\hat{H} = \sum_{pq} t_{pq} c^\dag_p c_q + \frac{1}{2}\sum_{pqrs} V_{pqrs} c^\dag_p c^\dag_q c_s c_r$, 
Using localized molecular orbitals, where each orbital is exponentially localized with radius $R_\text{loc}$, $t_{pq}$ is sparse.  
However, because of the Coulomb tail,  $V_{pqrs}$ is only diagonally sparse, with asymptotic long range contributions in $V_{pqpq}$, of which there are $O(L^2)$ terms.

More fundamentally, ${H}$ projected into the excitation space is \emph{not} $O(1)$ sparse, even for a local ${H}$. For example, even the tight-binding Hamiltonian for $L$-sites leads to $O(L)$ non-vanishing $\langle {}_i^a| {H} |0\rangle$ matrix elements in the localized molecular orbital basis (assuming $n_e = O(L)$) because every particle in the Fermi vacuum can do a single-particle hop. Such hops are vacuum fluctuations that change $m$, and to restore sparsity we have to use a formulation which avoids them. 

% In many-body physics language, these non-local terms arise from vacuum fluctuations, i.e. particles and holes can be created and destroyed out of the Fermi vacuum. To restore locality, we have to work within a theory where these types of vacuum fluctuations do not arise. 

We now explain the two resolutions mentioned in the main text more explicitly. The first is to restrict ourselves to fixed $m$-fold excitations only, in which case contributions from other spaces have to be folded in. This gives rise to the \emph{effective} Hamiltonian of the Bethe-Salpeter equation, the first post-mean-field method we study. The second possibility is to formulate the theory in terms of a different effective ``Hamiltonian'', the
Liouvillian superoperator ${L} = [{H}, \cdot]$. The commutator  ensures a connected structure; 
%to the theory; $[\hat{H}, \hat{O}]$ is localized to the same region of space as a local $\hat{O}$. 
its projection to the ordered excitation basis $\langle \mu |[{H}, \cdot]| \mu'\rangle$ 
%(where $\mu, \mu'$ are strings in the subspace of excitations of up to $m$-fold excitations) 
is  geometrically ``local'' and sparse in the sense that the two strings $\mu$, $\mu'$ can differ only by excitations in the region of space around $(\cdot)$. This Liouvillian formulation underlies the linearized coupled cluster theory that is the second post-mean-field method we discuss.}
 
\subsection{Multi-exciton Hamiltonian}
In the Tamm–Dancoff approximation (TDA) to the Bethe–Salpeter equation (BSE), a single-exciton ($m=1$) state is expanded in the basis of electron–hole excitations, 
\begin{equation}
    \ket{\psi_{SE}} = \sum_{ia} c_{ia} \ket{{}_i^a}.
\end{equation}
where $\ket{{}_i^a}$ denotes the configuration of an electron promoted from an occupied orbital $i$ to an unoccupied orbital $a$. The summation implicitly assumes summing over occupied and virtual spaces separately for $i$ and $a$. The equation for amplitudes follows an effective Hamiltonian $\mathbf{A}_{SE}$ \cite{si:blase2020bethe}, defined as 
% \chris{the correct index is  different from what we discussed in the meeting. I checked with Jiace again; he was looking at Eq. (19) in https://arxiv.org/pdf/2406.11122v1, which he thought was physicist's notation, but it isn't as shown in Eq. (24). The below one should be correct as verified by us}
{
\begin{equation}
\sum_{i'a'} (\mathbf{A}_{SE})_{ia}^{i'a'}~c_{i'a'}  = \sum_{a'} f_{aa'} c_{ia'}  - \sum_{i'} f_{ii'} c_{i'a} + \sum_{i'a'} (V_{ai'ia'} - W_{iaia'}) c_{i'a'}
\label{si:eq:exciton}
\end{equation}
}
with $V$ the bare Coulomb interaction and $W$ the statically screened Coulomb interaction. The first two terms describe single-particle contributions from the Fock operator for electrons and holes (and, if one uses GW-BSE within the quasiparticle approximation, the Fock operator also contains the (static) contribution of the electron self-energy), while the last term describes the electron–hole interaction. %If one incorporates spin, the kernel is replaced by $K_{iajb} = \alpha v_{iajb} - W_{ijab}$ where $\alpha = 2$ or $0$ depending on whether one targets singlet or triplet excited states, respectively.

Analogously, a biexciton ($m=2$) can be described as a correlated state of two excitons. In the TDA, the biexcitonic wavefunction is written as
\begin{equation}
\ket{\psi_{BE}} = \sum_{ia,jb} c_{ia,jb} \ket{{}_{ij}^{ab}}
\end{equation}
where $\ket{{}_{ij}^{ab}}$ denotes a doubly excited configuration with two holes $(i,j)$ and two electrons $(a,b)$. The amplitudes $c_{ia,jb}$ satisfy an effective Hamiltonian $\mathbf{A}_{BE}$ \cite{si:steinhoff2018biexciton} that naturally generalizes the single-exciton case, %\chris{same as above; see Eq. (5)}
{
\begin{align}
\sum_{i',j',a',b'} (\mathbf{A}_{BE})_{ia,jb}^{i'a',j'b'} c_{i' a', j' b'} = 
&\sum_{a'} f_{aa'} c_{ia',jb} - \sum_{i'} f_{ii'} c_{i'a,jb}
+\sum_{b'} f_{bb'} c_{ia,jb'}-\sum_{j'} f_{jj'} c_{ia,j'b} \nonumber \\
& +\sum_{a'b'} W_{a a' b' b}\, c_{i a', j b'}
      +\sum_{i'j'} W_{i i' j' j}\, c_{i'a, j' b} \nonumber \\
& +\sum_{i'a'} \left(V_{a i' i a'} - W_{i a i a'}\right) c_{i'a',jb}
      +\sum_{j'b'} \left(V_{b j' j b'} - W_{j b j b'}\right)\, c_{i a, j'b'} \nonumber \\
& +\sum_{a'j'} \left(V_{a j' j a'} - W_{j a j a'}\right)\, c_{i a', j' b}
      +\sum_{b'i'} \left(V_{b i' i b'} - W_{i b i b'}\right)\, c_{i' a, j b'} 
\end{align}}
where the first line again represents single-particle Fock contributions for each electron and hole. The second line accounts for electron–electron and hole–hole interactions (through $W$). The last two lines account for electron–hole interactions (through $V$ and $W$). This procedure can be generalized to cases for $m \ge 3$, by similarly describing electron-electron or hole-hole interactions through $W$ and electron-hole interactions through $V$ and $W$.

In a fixed local atomic basis, both the Coulomb and Fock matrix elements are bounded by $O(m^2)$ (which we treat as $O(1)$). Suppose the above operators are truncated to a sparsity $s \sim 2mR_c^D$, then the operator norm is bounded by $O(s)$, and the Frobenius norm is of order $O(\sqrt{s \times \text{matrix dimension}})$.

% which for fixed $m, D$, we regard as $O(1)$. Then, the eigenvalues 
% the operator norm is bounded by $O(2mR_c^D)$. This property allows for an efficient sparse block-encoding. {The eigenvalues of the operators are bounded by a subleading expression e.g. $\sim mD$}, thus the Frobenius norm is of order $O(\sqrt{\text{matrix dimension}})=O(\sqrt{L_c^{2mD}})$.

\subsection{Linearized coupled cluster}
The coupled cluster method assumes the exponential ansatz $\ket{\Psi} = e^T \ket{0}$, where $\ket{0}$ is a mean-field Slater determinant and ${T} = T_1 + T_2 + \ldots = \sum_\mu t_\mu {O}_\mu$. Here, ${O}_\mu$ excites to determinant $|\mu\rangle$, and $T_m$ contains the subset in the $\mu$ sum corresponding to $m$-fold excitations. Plugging the ansatz into the Schr\"odinger equation $H\ket{\Psi} = E\ket{\Psi}$ and inserting $e^{-T}$ to the left, one obtains $e^{-T} H e^T \ket{0} = E\ket{0}$. Projecting to $\ket{0}$ and the excited state manifold $\ket{\mu} = O_\mu \ket{0}$ results in the governing equation for $t$-amplitudes
\begin{align}
    &\bra{0}e^{-T} H e^T \ket{0} = E \\
    &\bra{\mu}e^{-T} H e^T \ket{0} = 0.
\end{align}
The linearized coupled cluster framework approximates $e^{-T} H e^T$ with the Baker-Campbell-Hausdorff expansion and truncates 
at the first order, yielding $e^{-T} H e^T \approx H + [H, T]$. This gives the equations
\begin{align}
    &\bra{0}(H + [H, T]) \ket{0} = E \label{si:eq:LCC_energy} \\
    &\bra{\mu}(H + [H, T]) \ket{0} = 0. \label{si:eq:LCC}
\end{align}
Inserting $T=\sum_\nu t_\nu O_\nu$, Eq.~\eqref{si:eq:LCC} becomes a linear equation
\begin{equation}
    \sum_\nu t_\nu \bra{\mu}[H, O_\nu] \ket{0} = -\bra{\mu} H \ket{0}.
\end{equation}
where the linear operator to be inverted is $\bra{\mu}[H, O_\nu] \ket{0}$ and $t_{\nu}$ is the solution. After solving for $t_{\nu}$, Eq.~\eqref{si:eq:LCC_energy} can be used to compute the correlation energy $E_c = E-E_0=\bra{0} H T_2 \ket{0}$ where $E_0 = \bra{0}H\ket{0}$. Here, we have used $\bra{0} TH\ket{0}=0$ and assumed Brillouin's theorem which states that $\bra{0} H T_1\ket{0} =0$.
%We briefly describe the nature of the matrix elements. The term $\bra{\mu}H\ket{\nu}$ is identical to that in configuration interaction and can be evaluated using the Slater–Condon rules \cite{si:helgaker2000second}. The contribution $\bra{\mu} O_\nu H \ket{0}$ is nonzero only if $\mu$ and $\nu$ correspond to the same excitation, giving $\bra{\mu}O_\nu H \ket{0}=E_0$, or if $\mu$ fully contains the excitations in $\nu$ and their excitation ranks differ by two (a difference of one is excluded due to Brillouin's theorem for a block-diagonal Fock matrix $f$).
{Each matrix element consists of at most $O(m^2)$ linear combinations of one- and two-electron integrals.} Similar to the multi-exciton Hamiltonian, the operator norm is $O(s)$ while the Frobenius norm is $O(\sqrt{s \times \text{matrix dimension}})$.

\section{Cost of quantum algorithms}

\subsection{Gate complexity of Hartree-Fock state preparation in the distinguishable particle basis}

As described in the multi-exciton section, we may wish to prepare a Hartree-Fock initial state. Using the procedure in Ref.~\cite{si:berry2018improved}, the gate complexity of the antisymmetrizer is {$m$ $\times$ polylog factors}, and the orbital rotations require $2m$ applications of unitaries of size $O(V_1) = O(L_c^D)$, both contributing additively to the total cost; thus the cost is negligible. 

\subsection{QROM versus QRAM in block-encoding}

In our block-encoding, we make extensive use of quantum read-only memory. For example, when loading the one-electron integrals, this allows us to access $\ket{i}\ket{j}\ket{0} \rightarrow \ket{i}\ket{j}\ket{f_{ij}}$ as an oracle (and similarly for two-electron integrals). The QROM can be used for the ``sparse-access oracles'' in sparse block-encoding schemes \cite{si:gilyen2019quantum,si:camps2022fable,si:camps2023explicit,si:sunderhauf2024blockencoding,si:lemieux2024quantum} (one needs separate oracles to query the positions of non-zero entries, which are easy to construct in our case). We can then use $O(1)$ queries of the oracles to construct the block-encoding.

The query cost of QROM is at most the size of the data \cite{si:babbush2018encoding}. This figures prominently in the ratio of quantum to classical costs. In some cost models, switching to QRAM
\cite{si:giovannetti2008quantum,si:aaronson2015read,si:arunachalam2015robustness,si:matteo2020faulttolerant,si:hann2021resilience,si:chen2023efficient} allows for a reduction of the query cost to logarithmic in the size of the data. However, there are subtle complications related to the cost model and error correction, thus we do not use QRAM here.

\subsection{Distinguishable particle representation}

In preparation of the canonical Hartree-Fock state, we mentioned the distinguishable particle representation. This is related to the excitation basis representation, but is a larger basis labelled by the integers $|ijk\ldots abc\ldots )$ without ordering. This  labels a first quantized basis with distinguishable particles and holes, and the correct fermionic state is given by $|ijk\ldots abc\rangle = \mathcal{A} |ijk \ldots abc)$, where $\mathcal{A}$ is the antisymmetrizer.

\subsection{Initial state overlap and preparation}

In the discussion of the multi-exciton problem, quantum algorithms to determine specific eigenvalues contain a $1/\gamma$ dependence where $\gamma$ is the initial state overlap. In Ref.~\cite{si:lee2023evaluating} which discussed the standard quantum simulation problem, it was argued that there is empirical evidence to suggest that for some chemical problems of interest, the success of widely used heuristics to prepare good initial states for quantum algorithms coincides with the existence of efficient classical heuristics that solve the standard quantum simulation problem with polynomial cost in system size. When this is true, it eliminates the exponential advantage in system size.

In the current setting, the classical cost depends on the correlation volume, and the exponential advantage is in $m \times D$. Since we do not scale $m, D$, asymptotic exponential advantage in $m$, $D$, is not part of our considerations. Nonetheless, if the exponential dependence on $m$ is fully removed in the classical method, we lose the main source of asymptotic polynomial speedup in $L_c$. Thus, should the analogous conjecture about the relationship between efficient state preparation (in relevant systems) and efficient classical heuristics (i.e. polynomial in $m$) hold true, we similarly eliminate asymptotic quantum advantage with correlation length.

This illustrates the difficulty of discussing asymptotic quantum advantage versus classical heuristics in the eigenstate setting. However, the correlation length $L_c$ is also not a natural problem parameter to scale: typically one is interested in solving a specific problem or set of problems, which have specific values for $L_c$. Thus our view is that rather than analyzing asymptotic advantage, one needs to establish advantage through a precise resource estimate. However, the theoretical asymptotic advantage (without taking into account $1/\gamma$) motivates the possibility of large practical advantage for eigenvalue estimation.

Alternatively, as motivated in the text, cleaner analysis of advantage can be obtained in the dynamical BSE setting or in the discussion of linearized coupled cluster theory, where the initial state overlap does not enter.

{
\subsection{Block-encoding schemes and total costs}

We follow approaches for sparse block-encoding similar to Section 6.5 of \cite{si:lin_lecture_2022}.
%Lemma 48 of \cite{si:gilyen2019quantum}
Suppose we wish to block-encode an $N \times N$ matrix $A_{IJ}$ with row- and column-sparsity $\leq s$, where $I$ and $J$ are indices corresponding to rows and columns, respectively. Assume we have the row- and column-sparse-index oracle
\begin{equation}
\begin{split}
    & O_{\mathrm{row}}: \ket{I}\ket{k} \rightarrow \ket{I} \ket{r_{Ik}}\\
    & O_{\mathrm{col}}: \ket{l}\ket{J} \rightarrow \ket{c_{lJ}} \ket{J}\\
\end{split}
\end{equation}
where $r_{Ik}$ and $c_{lJ}$ indicate the column- and row-index of the $k$- and $l$-th non-zero element of row $I$ and column $J$, respectively. Additionally, assume we have the amplitude oracle
\begin{equation}
    O_{\mathrm{amp}}: \ket{I}\ket{J}\ket{0} \rightarrow \ket{I} \ket{J} \left( A_{IJ} \ket{0}+ \sqrt{1-|A_{IJ}|^2} \ket{1} \right),
\end{equation}
then we can construct a block-encoding of $A$ using exactly one query to $O_r$, $O_c$ and $O_a$ and additional $O(\log(N))$ gates, with subnormalization $s$. Further optimization is possible with alternative definitions of the oracle \cite{si:camps2023explicit,si:sunderhauf2024blockencoding}, although the asymptotic scaling doesn't change. 

Adapted to our setting, we assume the index oracles can be constructed by efficiently iterating over the neighbors of the local orbitals. The construction of the amplitude oracle depends on the input models, which will be described below. In both the exciton and LCC cases, the effective Hamiltonian is a linear combination of $O(m^2)$ terms of $f$, $V$, and $W$, and can therefore be constructed via the addition of block-encodings of each of them ~\cite{si:gilyen2019quantum}. We focus on the block-encoding of $V$
\begin{equation}
    O_{V,\mathrm{amp}} : \ket{i}\ket{j}\ket{k}\ket{l}\ket{0}
    \;\rightarrow\;
    \ket{i}\ket{j}\ket{k}\ket{l}
    \left(
        V_{ijkl}\ket{0}
        +
        \sqrt{1 - \lvert V_{ijkl} \rvert^2}\ket{1}
    \right)
\end{equation}
and the procedures for $f$ and $W$ follow analogously. 

\underline{Integrals.} 
From a classical computation of integrals, we can access the entry oracle 
\begin{equation}
    O_{V,\mathrm{entry}} : \ket{i}\ket{j}\ket{k}\ket{l}\ket{0}
    \;\rightarrow\;
    \ket{i}\ket{j}\ket{k}\ket{l}
    \ket{V_{ijkl}}
\end{equation}
by loading through a QROM circuit, where $\ket{V_{ijkl}}$ is a $b$-bit description of $V_{ijkl}$. For simplicity, we assume this is exact. $O_{V,\mathrm{amp}}$ can be constructed from two queries of $O_{V,\mathrm{entry}}$ and additional $\text{polylog}(V,1/\epsilon)$ gates and ancillas, e.g. see work of Berry et al. \cite{si:berry2015hamiltonian}.

\underline{Atomic data.} We load the description of localized molecular orbitals. Techniques by Babbush et al.~\cite{si:babbush2017exponentially} can be adapted to construct the amplitude oracle, which computes a Riemann sum over a grid basis. 
Here we outline a potentially more efficient alternative approach. Each localized molecular orbital is expanded in $N_{\mathrm{aux}}$ basis functions,
\begin{equation}
    \varphi_i(\vec{r}) = \sum_{m=0}^{N_{\mathrm{aux}}-1} c_{i,m}\,\chi_{i,m}(\vec{r}),
\end{equation}
where $\chi_{i,m}(\vec{r})$ admits efficient evaluation of one- and two-electron integrals, e.g., Gaussian-type orbitals~\cite{si:helgaker2000atomic,si:helgaker2000gaussian}. The basis functions are accessed via the oracle
\begin{equation}
    O_{\chi} : \ket{i}\ket{m}\ket{0}
    \rightarrow
    \ket{i}\ket{m}\ket{\vec{\alpha}_{i,m}},
\end{equation}
where $i$ labels localized molecular orbitals, $m$ labels basis functions, and $\vec{\alpha}_{i,m}$ encodes $c_{i,m}$ together with the parameters of $\chi_{i,m}$ (e.g., angular momenta, center, and decay rates for primitive Cartesian Gaussian-type orbitals). Given $\vec{\alpha}_{i,m}$, one- and two-electron integrals can be evaluated in $\mathrm{polylog}(V,1/\epsilon)$ time using, for example, recurrence relations for Cartesian Gaussian-type orbitals~\cite{si:obara1986efficient,si:helgaker2000molecular}, which we denote through the oracle
\begin{equation}
    O_{\chi,\mathrm{int}}: \ket{\vec{\alpha}_{i,m_1}}\ket{\vec{\alpha}_{j,m_2}}\ket{\vec{\alpha}_{k,m_3}}\ket{\vec{\alpha}_{l,m_4}}\ket{0} \rightarrow \ket{\vec{\alpha}_{i,m_1}}\ket{\vec{\alpha}_{j,m_2}}\ket{\vec{\alpha}_{k,m_3}}\ket{\vec{\alpha}_{l,m_4}}\ket{I_{ijkl}^{m_1m_2m_3m_4}}.
\end{equation}
The amplitude oracle $O_{V,\mathrm{amp}}$ is then constructed by summing these integrals $V_{ijkl}=\sum_{m_1,m_2,m_3,m_4} I_{ijkl}^{m_1m_2m_3m_4}$ through a uniform superposition on the $\ket{m}$ registers, incurring an $O(N_{\mathrm{aux}}^4)$ overhead in subnormalization, where for exponentially localized orbitals $N_{\mathrm{aux}} = \text{polylog}(1/\epsilon)$.

\underline{Crystals.} In this setting, localized molecular orbitals are related by translation symmetry, so it suffices to load data only for orbitals within a single unit cell. For each unit cell with $N_{\text{loc}}=O(1)$ orbitals, let $0 \leq l \leq N_{\text{loc}}-1$ index an orbital within that cell. Similarly to before, we construct an oracle for the orbital parameters
\begin{equation}
    O_{\text{cell,}\chi} : \ket{l}\ket{m}\ket{0}
    \rightarrow
    \ket{l}\ket{m}\ket{\vec{\alpha}_{l,m}}.
\end{equation}
Additionally, we define an oracle
\begin{equation}
O_{\text{cell,ind}}:\ket{i}\ket{0}\rightarrow\ket{i}\ket{l},
\end{equation}
where $l$ is computed from the global orbital index $i$, which can be efficiently constructed with translation symmetry. We then use these oracles to compute the integrals coherently in the same manner as before, potentially using improved methods for periodic systems (e.g., \cite{si:sharma2021efficient}).

From the above, the total costs differ primarily by the type and amount of data loaded into quantum read-only memory (QROM) or improved variants \cite{si:babbush2018encoding,si:berry2019qubitization,si:low2024trading}. Based on the block-encoding schemes, we summarize the cost of each component in the entire algorithm in Table~\ref{si:tab:exciton_cost} for the exciton model and in Table~\ref{si:tab:LCC_cost} for the linearized coupled cluster model.

\section{Comparison with classical techniques}

The present work focuses on asymptotic scaling, rather than a system-specific benchmark against particular classical implementations. Some relevant classical techniques that have been used to accelerate similar electronic structure methods include tensor hypercontraction, stochastic tensor contraction, and quantics tensor trains.  A fully quantitative comparison would require fixing concrete model systems, truncation thresholds, and implementation details, which can significantly affect constant factors and crossover points. That said, structurally, the separation we identify arises from eliminating the explicit storage and update of the correlation volume. We briefly compare to the above classical approaches: (1) Tensor hypercontraction. THC reduces the rank of the underlying two-electron integral tensors. However, the effective Hilbert space dimension still scales exponentially with excitation rank $m$. Thus, while THC can lower constants and polynomial powers in system size, the exponential-in-$m$ growth of the excitation manifold remains. (2) Stochastic tensor contraction. Stochastic tensor contraction replaces deterministic tensor contractions with sampling. However, to avoid the sign-problem, efficient sampling requires storing and updating the excitation amplitude vector over the excitation manifold. The memory cost therefore, scales with the correlation volume, which grows exponentially in $m$. In contrast, our framework encodes the excitation manifold in $O(m \log L_c)$ qubits, achieving exponential compression in space relative to explicitly storing amplitudes. (3) Quantics tensor trains (QTT). QTT representations can avoid explicit exponential-in-$m$ scaling when the state admits low entanglement in the particle-coordinate representation. However, the entanglement structure of multi-exciton states -- especially in dynamical or high-energy regimes -- is not well characterized. We emphasize that entanglement in QTT is distinct from the usual sense of entanglement -- it characterizes how smooth or self-similar the function is in the first quantized representation. We do not expect such a property to hold for generic systems. Our framework does not rely on low entanglement structure and therefore avoids this potential bottleneck.

%For our case, consider $O(V)$ spin–orbital basis functions $\{\phi_j(\vec{z}_{\rho})\}$ for at $\vec{z}_{\rho}$, a position vector at grid point $\rho$. Our function oracle will then be 
%\begin{equation}
%    O_{\phi_j}: \ket{\rho} \ket{0}^{\otimes\log(M)} \rightarrow \ket{\rho} \ket{\tilde{\phi}_j(\vec{z}_{\rho})}
%\end{equation}
%where $\tilde{\phi}$ is the binary representation on $\log(M)$ bits. This oracle can be obtained from the classical computation of localized orbitals, and to implement it as a QRAM takes $\tilde{O}(V)$ time. Because orbitals are exponentially delocalized, integrals can be truncated to a region of size $\tilde{O}(1)$. Therefore, an amplitude code oracle with amplitude being 1- and 2-electron integrals can be computed on-the-fly at a cost of $\tilde{O}(V)$. In the LCC case, the excitation rank between $\mu$ and $\nu$ is at most $2$, therefore the matrix elements are a linear combination of $O(m^4)$ number of 1- and 2- electron integrals, thus the total cost stays as $\tilde{O}(V)$ to block-encode the operator.

\begin{table}[t]
    \centering
    \makebox[\textwidth][c]{
    {\begin{tabularx}{1.0\textwidth}{@{}|X|X|X|X|@{}}
    \hline
    Components & Integrals & Atomic data & Crystals \\
    \hline
    $C_{\mathrm{data}}$: Data-loading cost (QROM complexity) & $O(m^2 (dR_c)^{2D})$ & $\tilde{O}(m (dR_c)^{D})$ & $\tilde{O}(1)$ \\
    \hline
    $Q$: Queries to QROM in the block-encoding & \multicolumn{3}{c|}{$O(1)$}\\
    \hline
    $G_{\mathrm{BE}}$: Additional gates in the block-encoding &\multicolumn{3}{c|} {$\tilde{O}(1)$}  \\
    \hline
    $\alpha$: Subnormalization (sparsity of each term  $\times$ number of terms)  &\multicolumn{3}{c|} {$\tilde{O}(m^2 R_c^D)$} \\
    \hline
    $C_{\mathrm{init}}$: Initial state preparation (assuming localized) & \multicolumn{3}{c|}{$O(1)$} \\
    \hline
    $d$: Queries to block-encoding (matrix function) & \multicolumn{3}{c|}{$O(t+\log(1/\epsilon))$}\\
    \hline\hline
    $C_{\mathrm{BE}}$: Total block-encoding cost & $O(m^2 (dR_c)^{2D})$ & $\tilde{O}(m (dR_c)^{D})$ & $\tilde{O}(1)$  \\
    \hline
    $C_{\mathrm{total}}:$ Total cost & $\tilde{O}(m^4 d^{2D+1} R_c^{3D})$ & $\tilde{O}(m^3 d^{D+1} R_c^{2D})$ & $\tilde{O}(m^2 d R_c^{D})$\\
    \hline
     \end{tabularx}}}
    \caption{ Total costs for simulating exciton dynamics with a localized initial state. The block-encoding and total cost are computed by $C_{\mathrm{BE}} = C_\mathrm{data}Q + G_{\mathrm{BE}}$ and $C_{\mathrm{total}} = C_{\mathrm{init}}+C_{\mathrm{BE}}\alpha d$ respectively. Here, $\tilde{O}(.)$ ignores $\text{polylog}\left(V,1/\epsilon\right)$ factors. The first three rows correspond to components for block-encoding. For ground state preparation and ground energy estimation with initial state overlap $\gamma$ and energy error $\epsilon$, one encounters a multiplicative overhead of $O(1/\gamma)$ and $O(1/(\epsilon\gamma))$ respectively.}
    \label{si:tab:exciton_cost}
\end{table}

\begin{table}[t]
    \centering
    \makebox[\textwidth][c]{
    {\begin{tabularx}{1.0\textwidth}{@{}|X|X|X|X|@{}}
    \hline
    Components & Integrals & Atomic data & Crystals \\
    \hline
    $C_{\mathrm{data}}$: Data-loading cost (QROM complexity) & $O(V R_c^D)$ & $\tilde{O}(V)$ & $\tilde{O}(1)$ \\
    \hline
    $Q$: Queries to QROM in the block-encoding & \multicolumn{3}{c|}{$O(1)$}\\
    \hline
    $G_{\mathrm{BE}}$: Additional gates in the block-encoding &\multicolumn{3}{c|} {$ \tilde{O}(1)$}  \\
    \hline
    $\alpha$: Subnormalization (sparsity of each term  $\times$ number of terms) &\multicolumn{3}{c|} {$\tilde{O}(m^2 R_c^D)$}\\
    \hline
    $C_{\mathrm{init}}$: Initial state preparation  & $O(V R_c^D)$ & $\tilde{O}(V)$ & $\tilde{O}(1)$ \\
    \hline
    $d$: Queries to block-encoding (matrix function) & \multicolumn{3}{c|}{$O(\kappa \log(1/\epsilon))$}\\
    \hline\hline
    $C_{\mathrm{BE}}$: Total block-encoding cost & $O(V R_c^D)$ & $\tilde{O}(V)$ & $\tilde{O}(1)$\\
    \hline
    $C_{\mathrm{total}}:$ Total cost  & $\tilde{O}(m^2 V d R_c^{2D})$ & $\tilde{O}(m^2Vd R_c^D)$ & $\tilde{O}(m^2 d R_c^D)$\\
    \hline
      \end{tabularx}}}
    \caption{ Total costs for preparing $A^{-1}\ket{b}/\|A^{-1}\ket{b}\|$, with $A_{\mu\nu}=\bra{\mu}[H,O_{\nu}]\ket{0}$ and $b_\nu=-\bra{\nu}H\ket{0}$. The block-encoding and total cost are computed by $C_{\mathrm{BE}} = C_\mathrm{data}Q + G_{\mathrm{BE}}$ and $C_{\mathrm{total}} = C_{\mathrm{init}}+C_{\mathrm{BE}}\alpha d$ respectively. Here, $\tilde{O}(.)$ ignores $\text{polylog}\left(V,1/\epsilon\right)$ factors. The first three rows correspond to components for block-encoding. We assume the success amplitude is constant. To estimate the correlation energy per volume $E_c/V$, one incurs an extra $O(R_c^D/\epsilon)$ multiplicative cost, due to estimating the overlap with the reshaped state of $\bra{0}HT_2\ket{0}$ with norm $O(V R_c^D)$.}
    \label{si:tab:LCC_cost}
\end{table}
}

% REVTeX stores bibliography sizing metadata under a fixed label and replays
% front-matter notes in every bibliography. Give the SI bibliography its own
% metadata label and suppress the already-printed main-paper notes.
\begingroup
\makeatletter
\let\arxiv@origlabel\label
\let\arxiv@origref\ref
\def\arxiv@lastbib{LastBibItem}
\def\label#1{%
  \def\arxiv@arg{#1}%
  \ifx\arxiv@arg\arxiv@lastbib
    \arxiv@origlabel{SILastBibItem}%
  \else
    \arxiv@origlabel{#1}%
  \fi
}
\def\ref#1{%
  \def\arxiv@arg{#1}%
  \ifx\arxiv@arg\arxiv@lastbib
    \arxiv@origref{SILastBibItem}%
  \else
    \arxiv@origref{#1}%
  \fi
}
\let\@FMN@list\@empty
\makeatother

\endgroup
\end{document}